\documentclass[twocolumn,secnumarabic,amssymb, nobibnotes, aps, prd,superscriptaddress]{revtex4-2}

\usepackage{tikz}
\usepackage{here}
\usepackage{amsmath,amssymb}
\usepackage{bm}

\definecolor{blue}{RGB}{0 0 0}
\definecolor{green}{RGB}{0 0 0}

\begin{document}
	
	\title{Spatio-temporal analysis of electromagnetic field coherence in complex media}%
	
	\author{Thomas Fromenteze}%
	\affiliation{University of Limoges, CNRS, XLIM, UMR 7252, F-87000 Limoges, France}
	\author{Matthieu Davy}
	\affiliation{Univ Rennes,  CNRS,  IETR,  UMR6164,  F-35000  Rennes,  France}
	\author{Okan Yurduseven}
	\affiliation{Institute of Electronics, Communications and Information Technology (ECIT), Queen’s University Belfast, Belfast BT3 9DT, United Kingdom}
	\author{Yann Marie-Joseph}%
	\author{Cyril Decroze}%
	\affiliation{University of Limoges, CNRS, XLIM, UMR 7252, F-87000 Limoges, France}
	\email{corresponding author: thomas.fromenteze@unilim.fr}
	\date{September 2021}%
	
	\begin{abstract}
		We study the coherence in time and space of electromagnetic fields propagated through complex media. Whether for localization, imaging or telecommunication, the development of dedicated numerical techniques is generally based on the exploitation of simplified models considering either coherent or diffuse fields. The optimization of such applications in conditions of partial coherence can therefore be particularly challenging, requiring the development of hybrid algorithms adaptable to prior knowledge on the processed fields. The objective of this work is to provide numerical techniques for decomposing an electromagnetic field into subspaces that can then be filtered according to their level of spatial and temporal coherence. \textcolor{blue}{In contrast to the studies carried out on space-space transfer matrices notably used for the calculation of Wigner-Smith operators, these decompositions are carried out on space-time matrices in order to facilitate the study of temporal dispersion.} The theory is developed for illustrative purposes \textcolor{green}{using experimental results from} a leaky resonant system but seem to be applicable to any scattering and reverberating media capable of transforming localized and coherent excitations into complex and diffuse distributions. \textcolor{green}{To conclude this work, the proposed technique is exploited to improve image reconstruction in a millimeter-wave computational imaging demonstration. In the studied context and from a more general perspective, we propose a technique to select the most suitable subspaces for each application operating under conditions of partial coherence, whether these correspond in the most extreme cases to ballistic paths or to diffuse fields.} 
	\end{abstract}	
	
	\maketitle
	
	
	\section{Introduction}
	
	The study of electromagnetic wave propagation in complex media is motivated by both physical considerations and technological challenges involved in a growing number of practical applications~\cite{stephen1988temporal,vynck2021light,rotter2017light}. Whether one considers strongly scattering and/or reverberating environments, the excitation of spatially and temporally coherent waves within their boundaries shortly gives way to multimode regimes whose diversity tends to increase with dwell time~\cite{ishimaru1977theory}. Apart from integrable boundary conditions which can lead to periodic solutions, the more general complex and weakly symmetrical geometries of disordered propagation media are generally associated with chaotic or diffusive dynamics~\cite{kac1966can,stockmann1990quantum}. The sensitivity to initial conditions makes the long-term prediction of field distributions particularly delicate. The characterization of such media thus requires the definition of asymptotic and statistical properties such as the reverberation time~\cite{hill1994electronic} and the mean free path~\cite{savo2017observation}. In contexts where the analytical description of electromagnetic fields is notoriously challenging, it remains nevertheless necessary to promote operating conditions adapted to the constraints of each practical application. While free space propagation models can be considered in many cases for simplicity, the recent exploitation of new degrees of freedom offered by the diversity of complex propagation channels receives a growing interest from the scientific community. These efforts notably allow the optimization of telecommunication systems~\cite{jafar2007degrees,cadambe2008interference} and enable the focusing of energy~\cite{lerosey2007focusing,vellekoop2007focusing} and the localization of objects beyond the theoretical diffraction limits defined in free space~\cite{del2021deeply}. In associated contexts such as imaging through biological tissues, multiple scatterings are the source of limitations that impose the development of adapted processing techniques notably based on embedded detectors~\cite{popoff2010image,van2011scattering}, non-linear materials~\cite{hsieh2010imaging,de2016enhanced} or fluorescence~\cite{bertolotti2012non}.
	
	In this paper, we show that in open systems it is possible  to  discriminate  the  ballistic  and  diffuse  contributions  of  a  field  transmitted  through  a  complex  scattering  system  using  a  singular  value  decomposition  of  the frequency-space transfer matrix. We determine the level of temporal and spatial coherence of each subspace composing the studied field, facilitating the identification and extraction of independent structures, arranged from the first ballistic paths to the most complex and diffuse distributions.
	
	\textcolor{blue}{This work differs from most techniques developed for the study of complex media properties, which generally focus on the analysis of transmission matrices linking only spatial variables or the associated modes at the input and output of the considered systems~\cite{popoff2010measuring,rotter2011generating,ploschner2015seeing}. The temporal dimension is thus generally not considered, apart from couples of frequency samples allowing the calculation of partial derivatives for the evaluation of Wigner-Smith time-delay operators~\cite{smith1960lifetime,carpenter2015observation,del2021coherent}, facilitating the identification of eigenstates of scattering matrices~\cite{ambichl2017focusing,brandstotter2019shaping}. Focusing this work on the decomposition of transmission matrices linking input excitation points to the spatial and frequency dimensions of propagated fields, we propose here to study the evolution of coherence within complex media.} \textcolor{green}{Frequency-resolved measurements over large spectral bands thus provide an additional dimension to the transmission matrices, opening the way to new perspectives in the analysis of the propagation of electromagnetic fields in complex media. 		
	Significant advances have recently been published to reveal the information associated with this additional temporal dimension, in a context notably oriented towards the study of the modal diversity of optical fibers. Using swept-wavelength interferometric techniques, it has been demonstrated that it is possible to extract and amplify characteristic sub-spaces associated with propagation in studied complex media~\cite{carpenter2016complete,mounaix2019control}. In particular, the reference~\cite{mounaix2019control} has direct connections to the technique studied in this work, focusing on the eigenmodes of the covariance matrix computed from spatial transmission matrices determined at given propagation times. This work builds on previous contributions also based on singular value analysis and eigendecomposition, studying the characteristic spatial structures of matrices relating variables at the input and output of the studied media for the spatial focusing of waves~\cite{popoff2011exploiting} and the transmission of information through complex media~\cite{choi2011transmission}. In our work, we show that by adapting such formalisms to space-time transfer matrices, such spatial structures can be associated with temporal singular vectors, determining both the spatial and temporal coherence of the subspaces describing the propagation of waves through complex media.  A more detailed mathematical description of these references will be made after the definition of the formalisms associated with the proposed technique, allowing to better position our work in relation to the literature.}\\
	\indent Without loss of generality, we focus on the particular case of an electrically large metallic cavity coupled to the external world with radiating irises, motivated by computational microwave and millimeter-wave imaging applications presented in the last part of this study~\cite{hunt2013metamaterial,fromenteze2015computational,gollub2017large}. This work thus introduces methods for processing fields in complex environments using techniques that are as simple, efficient and intuitive as possible while remaining original, to the authors' knowledge. These efforts, however, fall within a context where singular value decompositions are applied in various research and engineering works such as blind source separation~\cite{de1994blind,gao2003blind}, the identification and extraction of acoustic and radar signatures~\cite{prada1996decomposition,pierri2006beyond}, the separation of diffracted waves in geophysics~\cite{freire1988application,shen2012seismic}, deep imaging in complex media~\cite{aubry2009random,lambert2020distortion,yoon2020deep} and speckle engineering in optics~\cite{devaud2021speckle}. \textcolor{blue}{This work could potentially find an application in the identification and study of regimes specific to disordered media such as localization and trapping phenomena characterized by a strong attenuation of the diffusion mechanisms~\cite{wiersma1997localization,leonetti2014observation,shi2018strong}. The extension of this work to the analysis and identification of particular regimes could then be explored by varying the local order level of the considered environments~\cite{zhu2020realizing,vynck2021light}.}\\
	\indent This paper is structured as follows: we first describe in the next section the considered cavity to facilitate the introduction of the coherence analysis techniques proposed in this work. The associated formalisms are developed in section 3, introducing coherence metrics and providing intuitive interpretations to the singular value decomposition. Section 4 presents a practical application of this work to the improvement of source localization experiments using frequency-diverse diffuse field radiation. The last section concludes this paper and discusses potential application perspectives.
	
	\section{Experimental conditions}
		
\begin{figure}[t]
		\centering
		\includegraphics[width=0.65\linewidth]{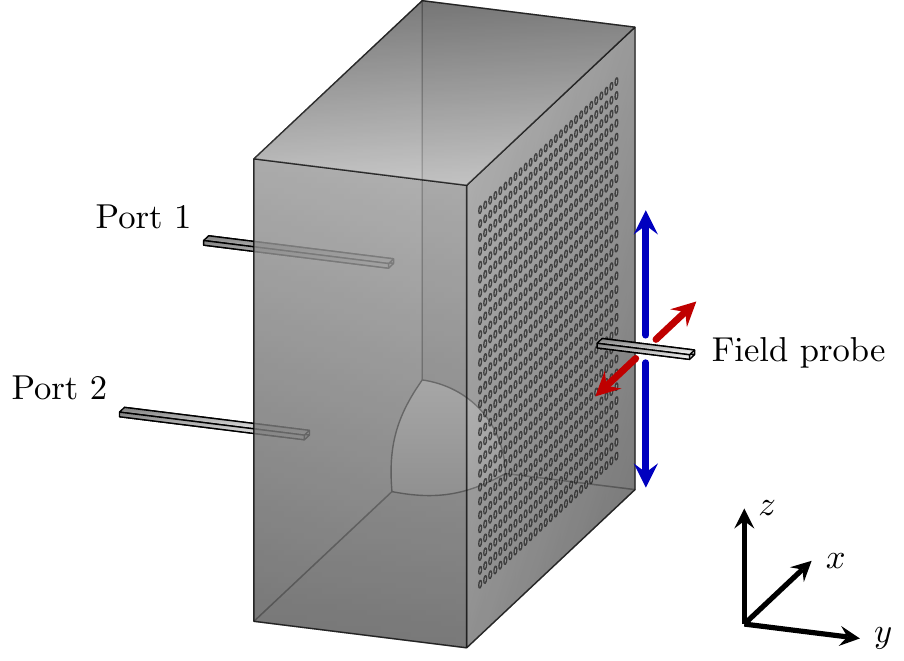}\\[0.2cm]
		\reflectbox{\includegraphics[width=0.65\linewidth]{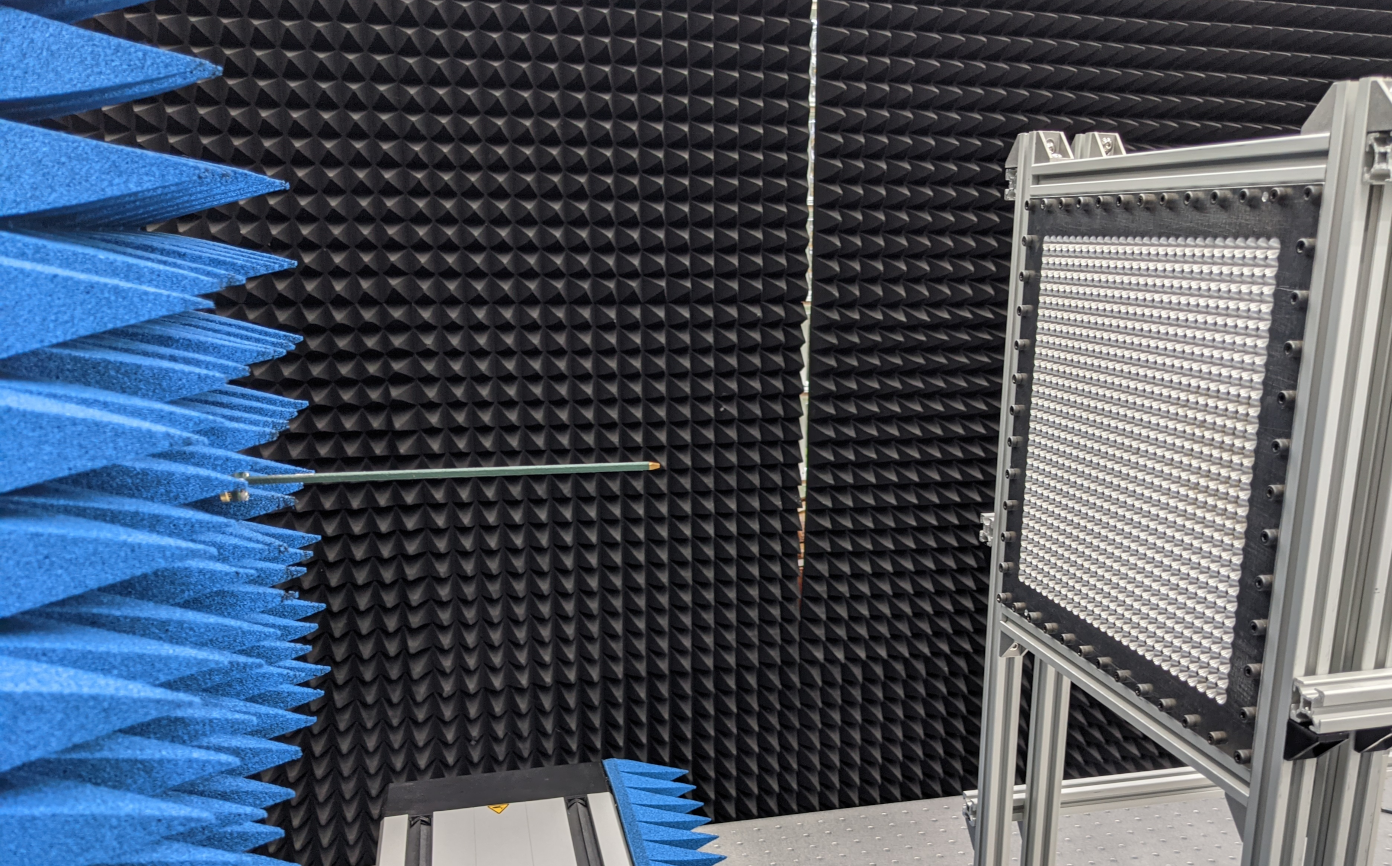}}
		\caption{Near field scan of a leaky reverberant medium. This aluminum cavity is excited at two locations by z-polarized WR-10 waveguides. Circular slits on the front panel radiate fields measured by a single-polarized scanning probe, oriented according to one of the two transverse polarizations $x$ and~$z$.}
		\label{fig:Cavity}
	\end{figure}

	In this work, we investigate the temporal and spatial evolution of an electromagnetic field bouncing into a metallic cavity. \textcolor{green}{All the measurements presented in this article are thus derived from data captured in an experimental context.} Excitation ports allow the injection of waves at two locations, which after a certain number of reflections are coupled towards the external world by the intermediary of irises (Fig.~\ref{fig:Cavity}). The latter is realized with polished aluminum panels, forming a parallelepiped whose internal dimensions are $251 \text{mm} \times 66 \text{mm} \times 251 \text{mm}$. The excitation is achieved with two WR-10 guide ports electrically polarized along the $z$ axis, with dimensions $2.54 \text{mm} \times 1.27 \text{mm}$. These are arranged on the rear face according to the arbitrarily chosen positions $(x,z) = (-37.52 \text{mm},-39.54\text{mm})$ and $(x,z) = (76.63 \text{mm},0\text{mm})$. An eighth of an aluminum sphere of 40mm radius is placed in one of the corners to limit the symmetries, ensuring a more homogeneous spatial distribution of the modes by combining regular and convex boundary conditions. The front face is perforated by a uniform array of $28 \times 28$ circular irises of 3mm diameter arranged along the $x$ and $z$ axes every 8mm. A 45$^\circ$ milling is performed for all the irises on the 5mm thickness of the front face. The radiation of this assembly is characterized by means of a near field scanner, moving a mono-polarized probe in a plane located at 14mm from the radiating surface. The characterization is carried out in the millimeter band from 70GHz to 100GHz, sampled by $n_{\nu}=$4001 measurement points. The scan is performed on a surface of $300\text{mm} \times 300\text{mm}$, spatially sampled by 201 points per side corresponding to a total number of $n_r$=40401 scanned positions. The space $\mathbf{r}$ swept by the field probe is discretized according to a sampling constraint of 1.5mm per transverse axis imposed by the highest frequency of the $\nu$ vector, corresponding to half a wavelength at 100GHz. The electric near field $E_{i}^{(p)} (\mathbf{r},\nu)$ radiated by the front panel is thus determined in a plane parallel to the latter according to one of the two transverse polarizations $i=(x,z)$ \textcolor{green}{and input port $p = (1,2)$}. \textcolor{green}{Four days were necessary for the complete characterization of the radiation of this cavity, performing intermediate verifications of the measured data for each excitation port and each transverse polarization. Due to the sequential Cartesian positioning of the field probe, these characterization times are necessarily much longer than those involving the use of spatial light modulators at optical wavelengths. As a comparison, the measurement of a transfer matrix determining the interaction between 254 input spatial modes, 650 output spatial modes for 1273 wavelengths is achieved in a little more than one hour in~\cite{mounaix2019control}.}

\begin{figure}[ht]
		\includegraphics[width=\linewidth]{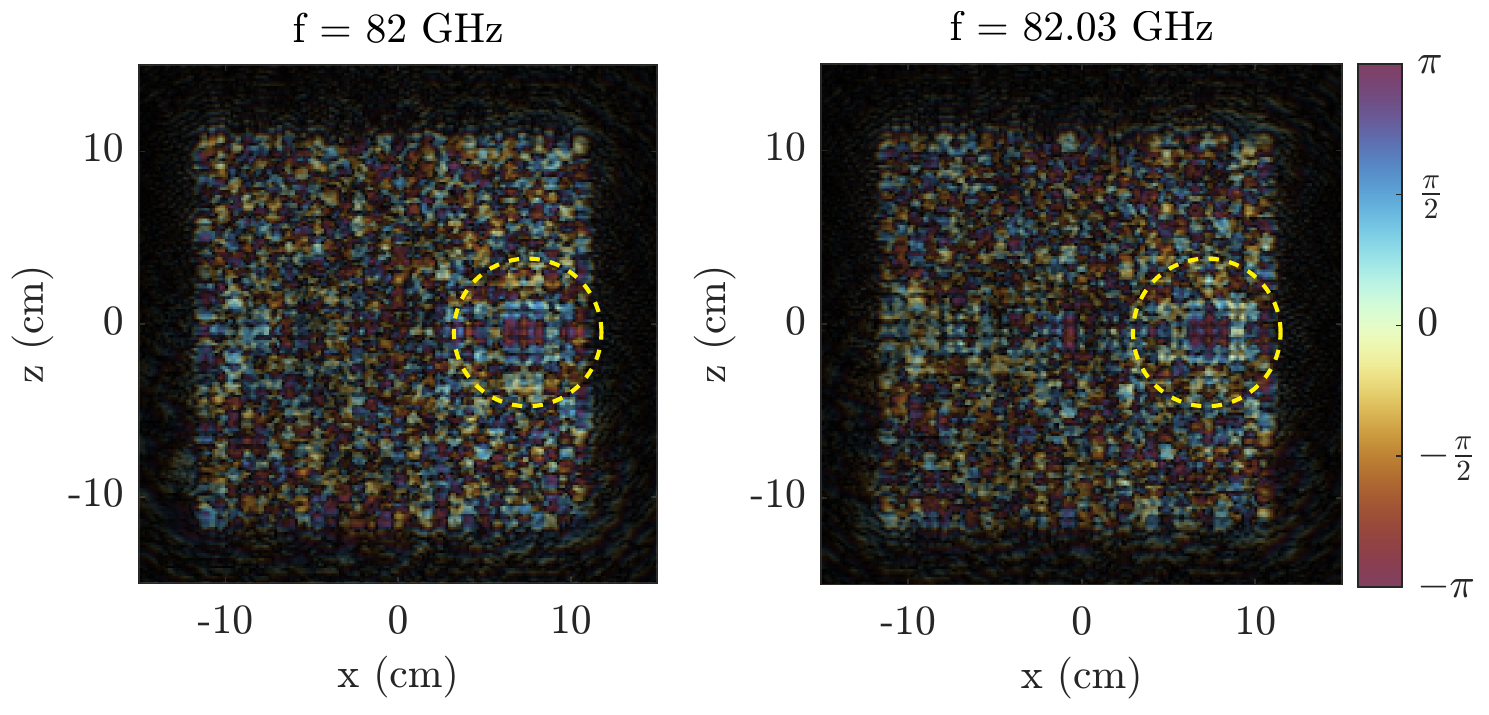}\\[-0.1cm]
		\caption{Electric fields $E^{(1)}_{z} (\mathbf{r},\nu)$ measured at 14 mm from the radiating aperture following the excitation of port 1. The phase (in radians) is represented in color and the opacity corresponds to the amplitude. \textcolor{green}{The field region surrounded by a dashed line has a spatial structure common to several frequency scans and appears to be related to a ballistic path.} The field distributions represented in this study are based on scientific colormaps optimized for color perception impairments~\cite{crameri2018scientific}.}
		\label{fig:Field}
	\end{figure}

	The evolution of a temporally and spatially localized excitation within such a medium tends to generate an increasingly complex field distribution, coupled to the external environment through the circular irises. The near field thus includes components reflecting a direct, strong and coherent coupling with the scanning probe as well as a succession of diffuse contributions, linked to an accumulation of wavefronts reflected in this reverberant environment. These characteristics are directly visible in the measured field at each frequency~(Fig.~\ref{fig:Field}). 
	
	In the given examples, the direct and co-polarized coupling between the cavity excitation port and the measurement probe forms a zone where the phase varies more weakly, around the location of the first port $(x,z) = (76.63 \text{mm},0\text{mm})$. The speckle pattern around this area changes rapidly with the analysis frequency and corresponds to random magnitude and phase distributions associated with the superposition of a large number of incident wavefronts. The electromagnetic field in perfectly chaotic cavities is indeed statistically isotropic, uniform and depolarized with universal statistics~\cite{gros2016lossy}.
	
	Having defined the experimental conditions in which the electric field is measured in space at each frequency of an operating band, a technique allowing the extraction of ordered subspaces according to their level of spatial and temporal coherence is presented in the next section.
	
	\section{Subspace decomposition by level of spatial and temporal coherence}
	
\begin{figure}[ht]
		\centering
		\includegraphics[width=1\linewidth]{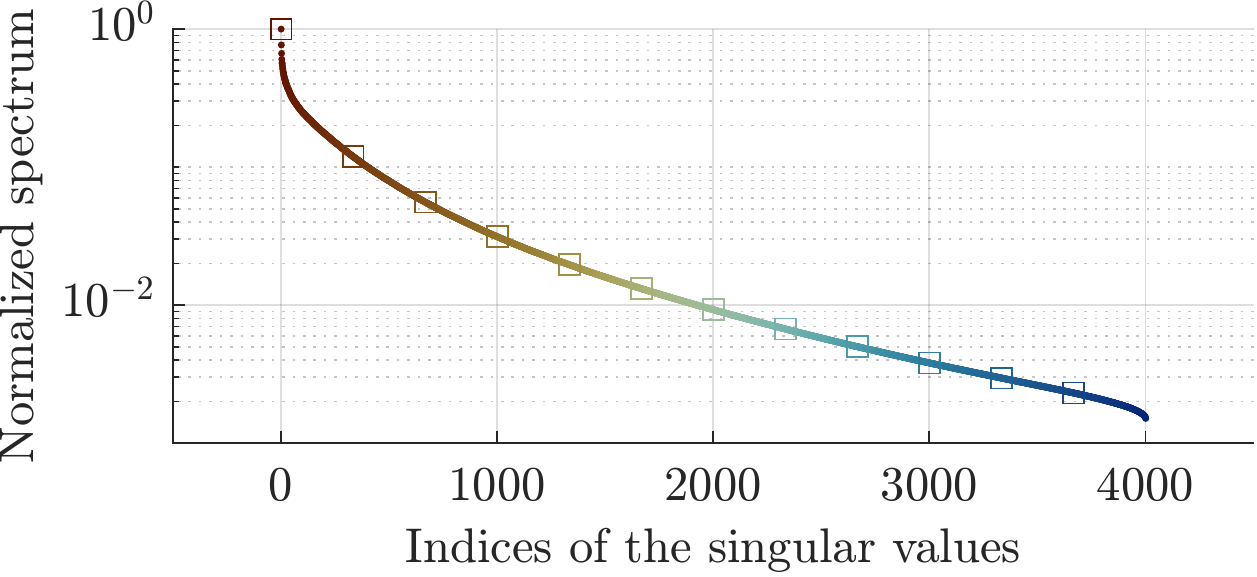}\\[-0.1cm]
		\includegraphics[width=1\linewidth]{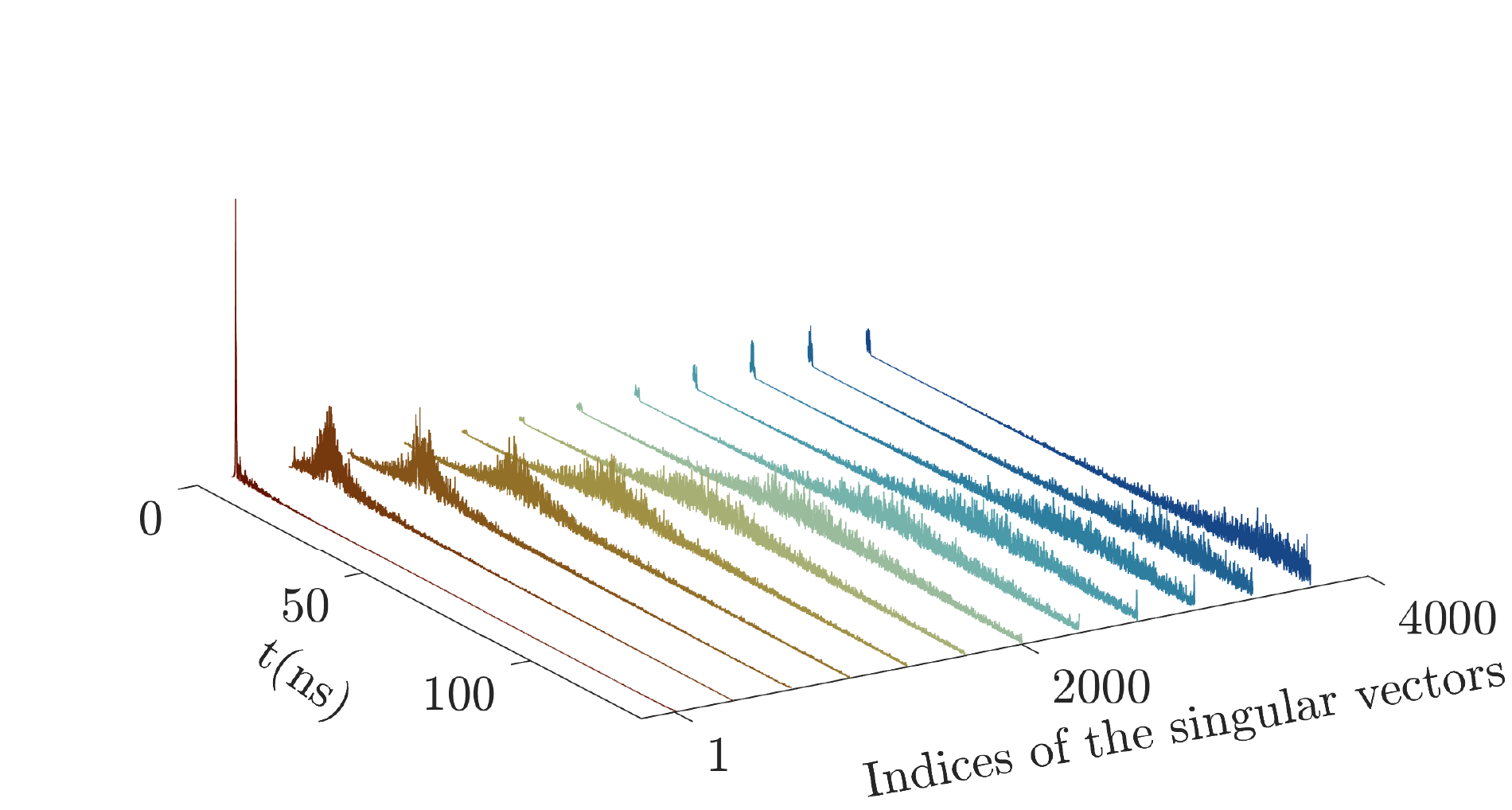}\\[-0.2cm]
		\caption{Top: Spectrum of singular values, highlighting the particularly significant contributions of a set of orthogonal subspaces associated with the first singular values. Bottom: Singular vectors $\bm{\mathfrak{u}}_n$ corresponding to the values highlighted by square markers.}
		\label{fig:SVD}
	\end{figure}

	In the given context, the singular value decomposition helps highlighting the predominant independent structures characterizing a measured field. We define a first complex matrix $\mathbf{E}_z^{(1)} \in \mathbb{C}^{n_\nu \times n_r}$, with $n_\nu$ and $n_r$ the respective numbers of elements of $\nu$ and $\mathbf{r}$. \textcolor{green}{Throughout this paper, we use bold notation for matrices and vectors which are distinguished with upper and lower case letters respectively.} In the first part of this study, we omit the systematic writing of the index of the excitation port~$(1)$ and of the polarization~$z$ to lighten the notations of the next developments applied in an identical manner to all the near fields studied thereafter. The singular value decomposition is defined as follows: 
	
	\begin{align}
	\mathbf{E} = \sum_n \sigma_n \mathbf{u}_n \mathbf{v}_n^\dagger 
	\end{align}
	
	\noindent where $\mathbf{u}_n \in \mathbb{C}^{n_\nu \times 1}$ and $\mathbf{v}_n \in \mathbb{C}^{n_r \times 1}$ are the singular vectors whose outer product forms an orthonormal basis, weighted by each associated singular value $\sigma_n \in \mathbb{R}$. The symbol $.^\dagger$ represents the conjugate-transpose operator. The near field matrix is thus decomposed into a set of orthogonal structures, sorted by amplitude contribution of the singular values and their associated vectors. This factorization thus represents an algebraic tool revealing the effective rank of a field matrix, exploited notably to define diversity metrics \cite{aubry2010singular,fuchs2019reduced,hoang2021spatial}. We note at this stage that the $\mathbf{u}_n$ vectors allow the definition of the frequency dimension of the field matrix to be characterized, thus revealing the connection between the main spatial field structures of $\mathbf{v}_n$ and the associated frequencies. In order to analyze the evolution of these field distributions in the time domain and to identify the associated level of temporal coherence, an inverse Fourier transform, denoted by $\mathfrak{F}^{-1}$, is applied to these singular vectors as $\mathfrak{u}_n(t) = \mathfrak{F}^{-1}(u_n(\nu))$. The attention of the readers must be drawn to the notation of the temporal vectors $\mathbf{\mathfrak{u}}_n$, stylized in order to limit a possible confusion with their frequency analogues $\mathbf{u}_n$. It seems worthwhile to point out that the singular values $\sigma_n$ and singular vectors $\mathbf{v}_n$ are identical whether a captured field is expressed in the frequency domain or the time domain. This property is justified insofar as the $\mathbf{v}_n$ vectors represent only spatial field structures and as unitary matrices preserve norms, thus not impacting the magnitude of the singular values. This invariance can also be illustrated by applying a discrete inverse Fourier transform matrix $\mathbf{W^{-1}}$ to the frequency domain field $\mathbf{E}$:
	\begin{align}
	\mathbf{E}_t
	&= \mathbf{W}^{\dagger} \mathbf{E}\\
	&= \mathbf{W}^{\dagger} \mathbf{U} \mathbf{S} \mathbf{V}^\dagger\\
	&= \mathbf{U}_t \mathbf{S} \mathbf{V}^\dagger
	\end{align}
	
	\noindent where $\mathbf{E}_t$ corresponds to the temporal field matrix at the output of the considered complex medium. The unitary matrices $\mathbf{U}$ and $\mathbf{V}$ are respectively composed of the singular vectors $\mathbf{u}_n$ and $\mathbf{v}_n$. The diagonal matrix $\mathbf{S} = \text{diag}(\bm \sigma)$ is constituted by the singular values arranged in descending order. Finally, $\mathbf{U}_t = \mathbf{W}^{-1} \mathbf{U}$ corresponds to the matrix of temporal singular vectors $\bm{\mathfrak{u}}_n$. These transformations finally allow the analysis of the evolution of the temporal singular vectors according to the level of the associated singular values (Fig. \ref{fig:SVD}). 
	
	The analyzed electric field is dominated by first components with limited temporal spread, corresponding to strong wavefronts that have weakly interacted with the considered medium. One can notice that the temporal coherence of the singular vectors then decreases rapidly, transiting from high amplitude pulses to diffuse distributions whose spread and average propagation time increase with the singular vector index, studied more carefully in the next part of this paper. 
	
\begin{figure}[ht]
		\includegraphics[width=\linewidth]{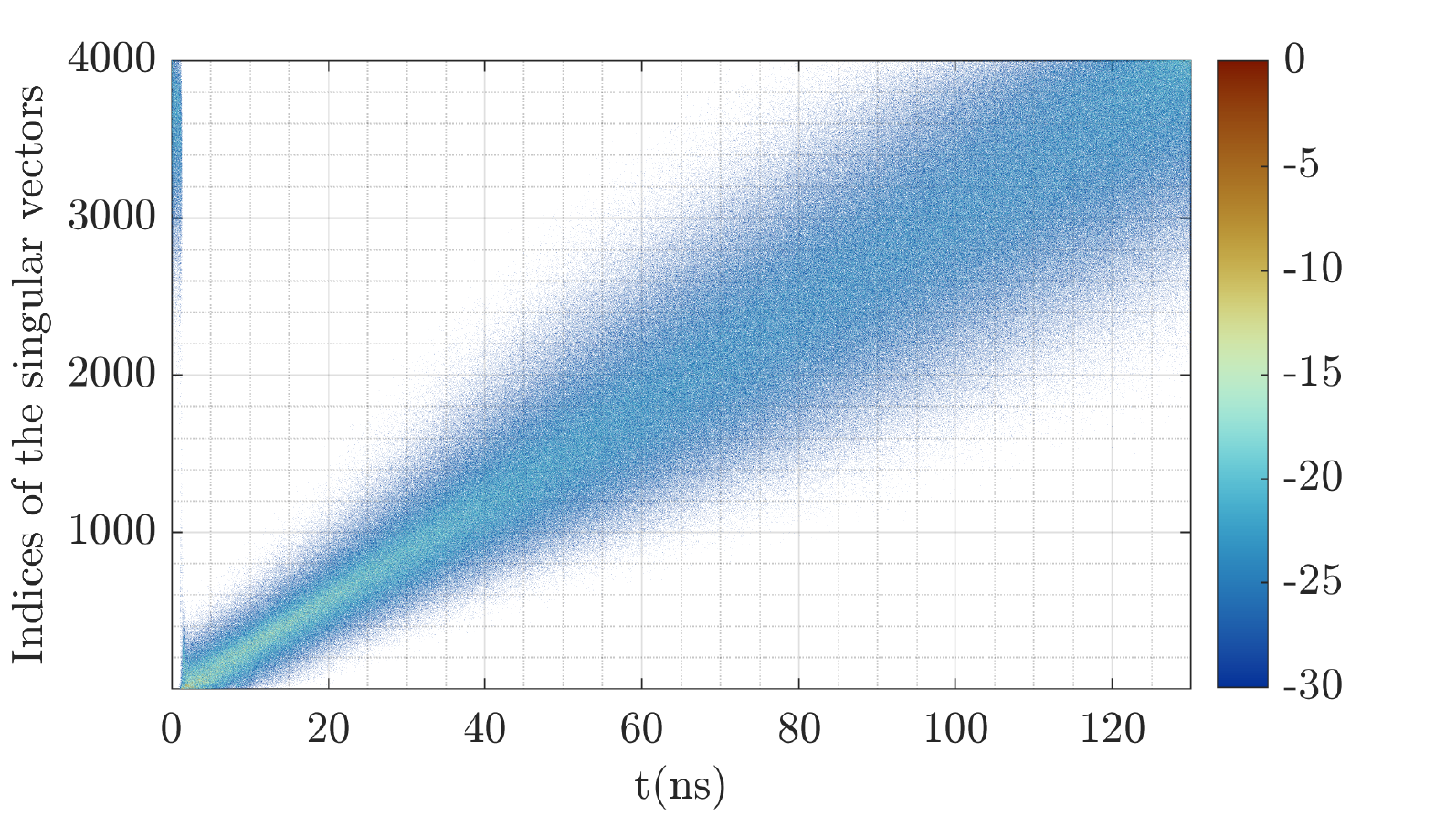}
		\caption{Matrix $\mathbf{U}_t$ of temporal singular vectors $\bm{\mathfrak{u}}_n$ highlighting the dwell time of the different associated spatial vectors $\mathbf{v}_n$ as well as the increasing modal diversity within this complex environment. The matrix is normalized and its magnitude is displayed in decibels.}
		\label{fig:Ut}
	\end{figure}

	The previous developments have highlighted the existing links between the temporal and spatial singular vectors, justifying that the matrix $\mathbf{U}_t$ allows the identification of the quantity of independent spatial structures necessary to characterize the field at each instant $t$ (Fig.~\ref{fig:Ut}). 
	
\begin{figure}[ht]
		\includegraphics[width=\linewidth]{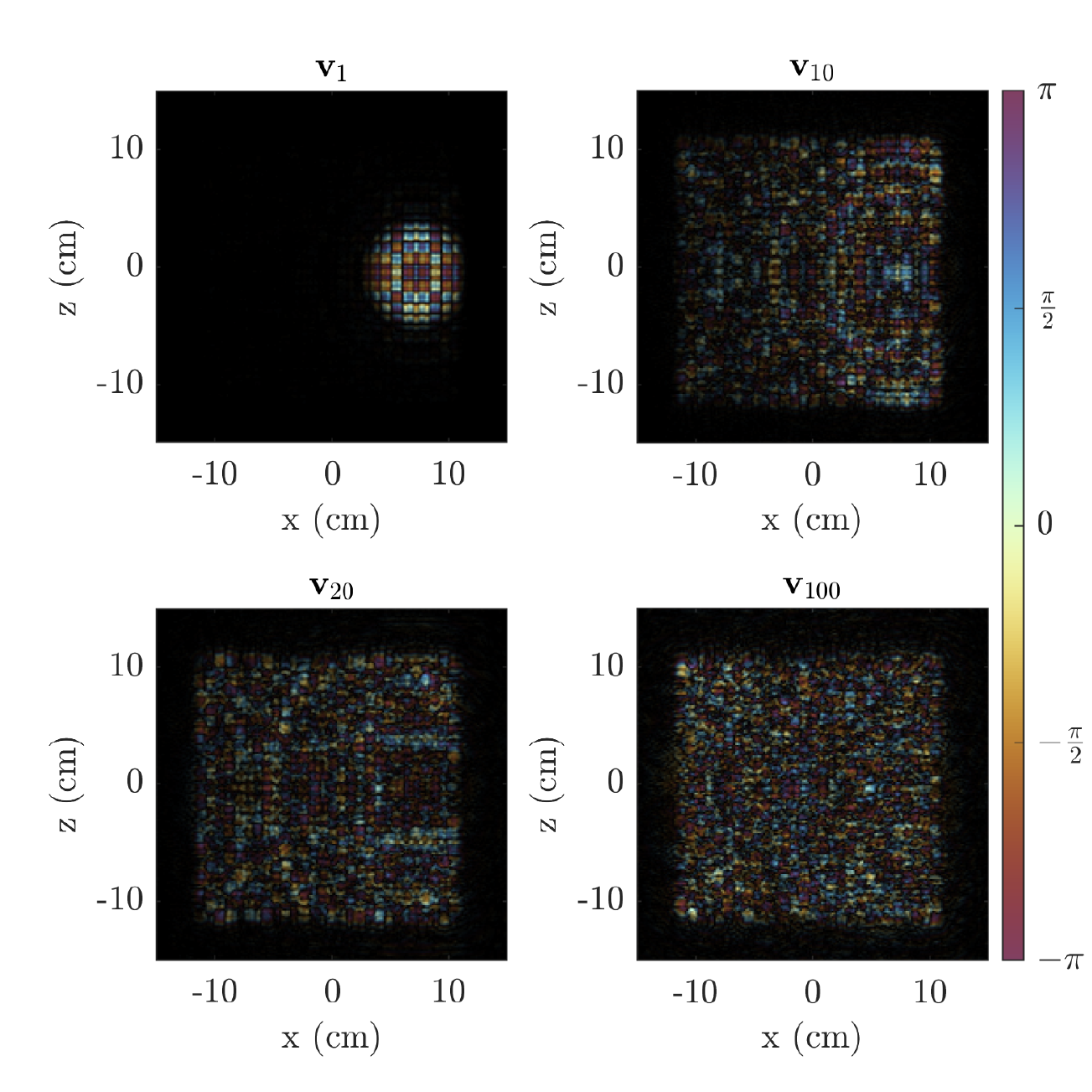}
		\caption{Four singular vectors $\mathbf{V}_n$ reshaped according to the two spatial dimensions $x$ and $z$. The phase (in radians) is represented in color and the opacity is defined by the magnitude.}
		\label{fig:SpatialModes}
	\end{figure}

	Such a simple visualization thus reveals crucial information on the temporal evolution of the spatial diversity of the field propagating within this environment, corresponding to the columns of this matrix where each sample is associated to a spatial singular vector. The distribution of values within this matrix also reveals the evolution of the dwell time of spatial structures associated with the same indices. Counter-intuitively, we observe that the first instants are associated with space vectors $\mathbf{v}_n$ whose indices are particularly high. Corresponding to propagation times lower than required to reach the front panel of the cavity, these non-zero values seem to be caused by a temporal aliasing phenomenon induced by the inverse Fourier transform associated with an incomplete damping of the impulse responses of this medium. An in-depth analysis is proposed by studying the distribution of spatial singular vectors $\mathbf{v}_n$, de-vectorized according to the two dimensions $x$ and $z$ to form associated matrices noted $\mathbf{V}_n$ \textcolor{green}{ and defined as $\mathbf{v}_n = \text{vec}(\mathbf{V}_n)$} (Fig.~\ref{fig:SpatialModes}). 
	
	These analyses depict the progressive increase in complexity of the measured field distributions in space. The growing number of independent structures required for their characterization represents as many degrees of freedom exploited by various applications of communication, localization and imaging.\\
	\color{green}
	\indent It seems meaningful at this point to emphasize more directly the link between our work and previous papers mentioned in the introductory section~\cite{popoff2011exploiting,choi2011transmission,mounaix2019control}. These references exploit decompositions of transmission matrices noted here $\mathbf{T} \in \mathbb{C}^{N_\text{out} \times N_\text{in}}$, linking $N_\text{in}$ spatial variables at the input and $N_\text{out}$ at the output of a complex propagation medium, determined for a given wavelength or time. These studies notably focus on the eigenvalue decomposition of $\mathbf{T}^\dagger \mathbf{T} \in \mathbb{C}^{N_\text{in} \times N_\text{in}}$, where $\dagger$ corresponds to the transpose-conjugate operator, revealing the associated principal spatial field structures. By realizing a singular value decomposition such that \mbox{$\mathbf{T}$ = $\mathbf{U}_T\, \mathbf{S}_T \mathbf{V}_T^\dagger$} and exploiting the properties of the unitary matrices $\mathbf{U}_T$ and $\mathbf{V}_T$, we have:
	\begin{align}
		\mathbf{T}^\dagger \mathbf{T} &= \mathbf{V}_T\, \mathbf{S}_T^\dagger \mathbf{U}_T^\dagger \mathbf{U}_T\, \mathbf{S}_T^\dagger  \mathbf{V}_T^\dagger \\
		&= \mathbf{V}_T\,   \mathbf{S}_T^\dagger\, \mathbf{S}_T\,  \mathbf{V}_T^\dagger
	\end{align}
	The eigenvectors $\mathbf{V}_T$ of $\mathbf{T}^\dagger \mathbf{T}$ then also correspond to the spatial singular vectors at the system input et les valeurs propres sont équivalentes au module carré des valeurs singulières. The singular vectors gathered in $\mathbf{U}_T$ correspond to the spatial structures characterizing the field at the output of the medium, orthogonal to each other according to the properties of unitary matrices. The injection of a singular vector $\mathbf{v}_{T,k}$ at the input of the system, corresponding to the singular value $\sigma_{k,T}$ extracted from the diagonal of the matrix $\mathbf{S}_T$, allows one to obtain at the output a field distribution $\hat{\mathbf{u}}_{T,k}$ such as:
	\begin{align}
		\hat{\mathbf{u}}_{T,k} &= \mathbf{T}\, \mathbf{v}_{T,k}\\
 					   &= \mathbf{U}_T\, \mathbf{S}_T \mathbf{V}_T^\dagger\, \mathbf{v}_{T,k}\\
	         		   &= \sigma_{T,k}\, \mathbf{u}_{T,k} 	
	\end{align}
	
	A measurement of the field distribution at the system output allows the retrieval of the singular vector $\mathbf{u}_{T,k}$ weighted by the associated singular value. In this context, these studies generally focused on the identification of the most significant singular values or eigenvalues, allowing to excite the most significant field structures. There does not seem to be a direct correspondence between our spatial singular vectors and those studied in these earlier works because the decompositions are performed here on space-time matrices. However, we find common characteristics associated to the largest singular values being exploited for the excitation of quasi-ballistic subspaces. \\
	\color{black}
	
	\textcolor{green}{We now proceed to analyze the results associated with our model.}	Having highlighted the link between the spatial and temporal characteristics of the field under consideration, it now seems useful to define metrics allowing the determination of the level of coherence of the associated subspaces. The temporal singular vectors $\mathbf{\mathfrak{u}}_n$ form wave packets localized around average propagation times $\bar{t}_n$ which increases with the index $n$ of the associated singular values. The latter can be determined by calculating the first temporal moment of each vector, also referred in the literature as the time centroid~\cite{lubcke1959frequenzbewertung,bauer1979time,sebbah1999fluctuations}:
	\begin{align}
	\bar{t}_n = \frac{ \int_{0}^{+\infty} t\, \lvert \mathfrak{u}_n(t)\lvert ^2 dt }{ \int_{0}^{+\infty} \lvert \mathfrak{u}_n(t)\lvert ^2 dt }
	\end{align}
	
\begin{figure}[h]
		\centering
		\includegraphics[width=0.95\linewidth]{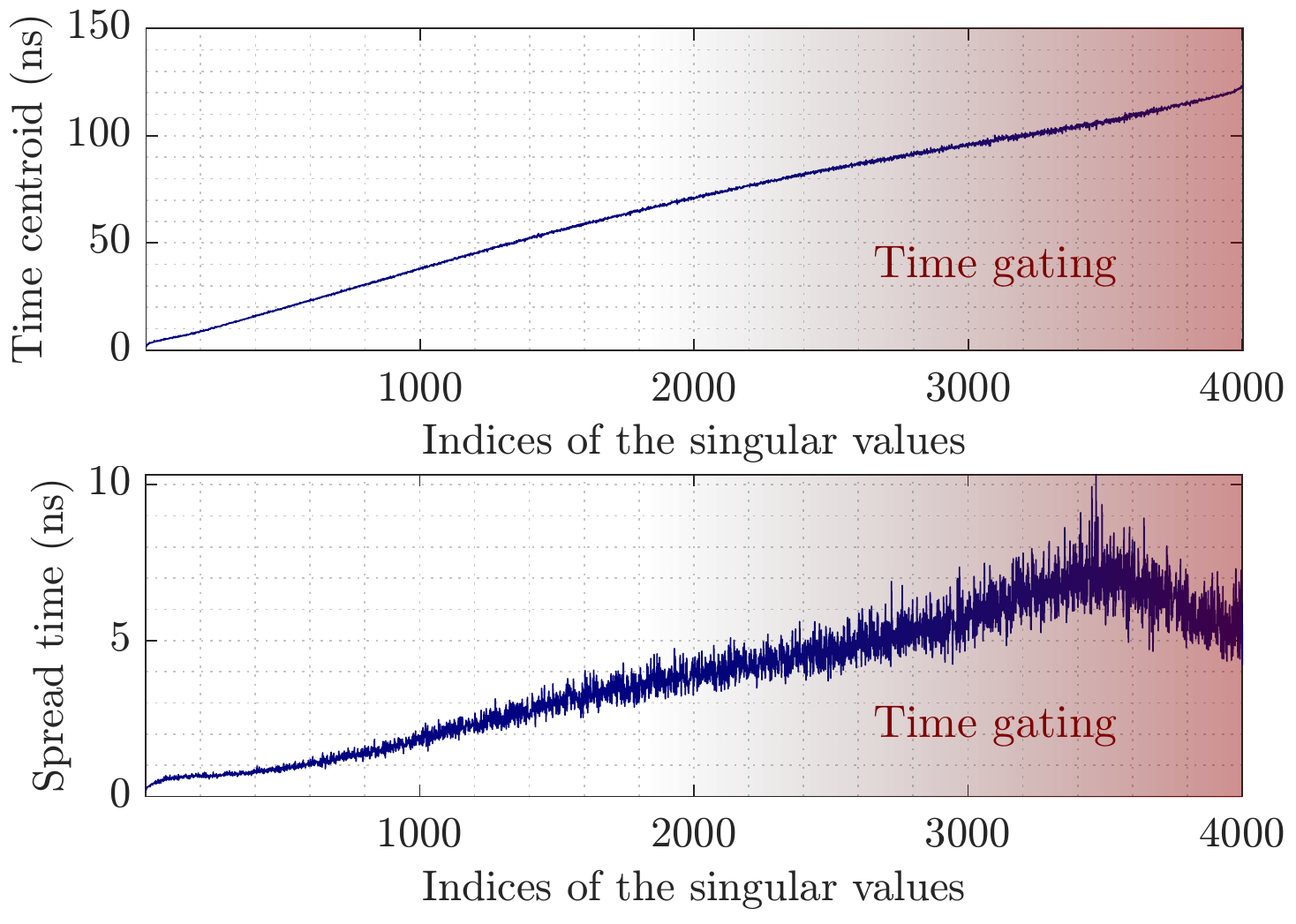}
		\caption{Time centroids $\bar{t}_n$ (top) and spread times $\tau_n$ (bottom) of the singular vectors $\mathbf{\mathfrak{u}}_n$.}
		\label{fig:centroid_spread}
	\end{figure}

	The results shown in Fig.~\ref{fig:centroid_spread} show quasi-linear dependency on the time centroid of each singular vector according to the associated indices.
	
\begin{figure}
		\centering
		\includegraphics[width=0.95\linewidth]{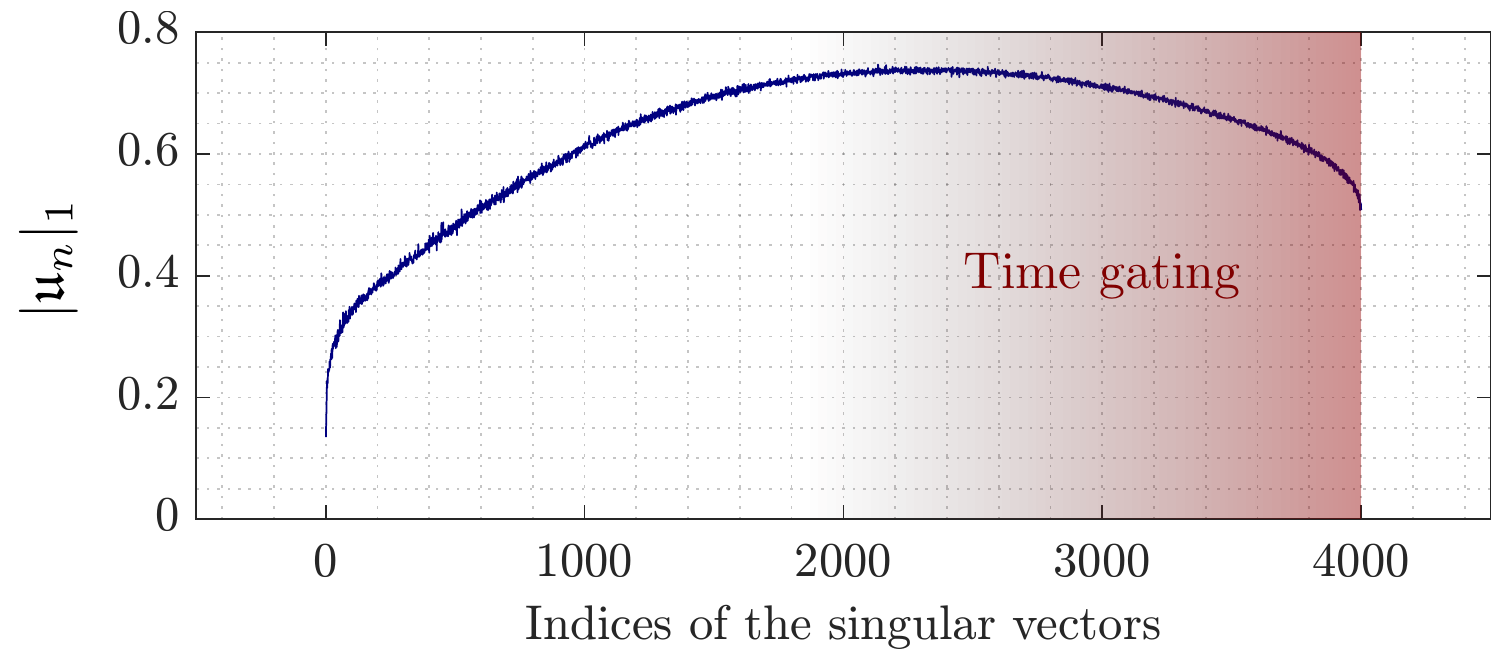}
		\caption{Evolution of the $\ell_1$ norm of the $\mathbf{\mathfrak{u}}_n$ vectors, defined as a temporal coherence metric.\vspace{-0.5cm}}
		\label{fig:utl1}
	\end{figure}

	Considering a physical optics approximation to the propagation within this reverberant environment, we deduce that the set of rays received at a given time $\bar{t}_n$ travels on average a distance corresponding to $c\, \bar{t}_n$, where $c$ corresponds to the speed of light. An average propagation time of 50 ns corresponds to a distance of about 15m, representing several tens of reflections within this medium and 5000 times the smallest wavelength considered for this study. The only information of spread time of each singular vectors $\mathbf{\mathfrak{u}}_n$ seems sufficient to determine the level of temporal coherence of the associated subspace. The extraction of these time constants can be achieved by considering exponential decay models related to the energy leakage from the studied domain at each time step~\cite{holloway2012early,hak2012measuring}, involving envelope variations of the form $\exp(-t/\tau_n)$ in intensity, or $\exp(-t/(2 \tau_n))$ in magnitude of the associated fields. This operation can be quite tedious and imprecise, achieving linear regressions on weighted random distributions expressed on a logarithmic scale. Adapted from the acoustic field, the backward integration method initially developed by Schroeder helps with the determination of reverberation times~\cite{schroeder1965new}. He astutely considered the following identity for this evaluation:
	\begin{align}
	\int_{0}^{t} \lvert \mathfrak{u}_n(t^\prime)\lvert ^2 dt^\prime = \int_{0}^{+\infty} \lvert \mathfrak{u}_n(t^\prime)\lvert ^2 dt^\prime - \int_{t}^{+\infty} \lvert \mathfrak{u}_n(t^\prime)\lvert^2 dt^\prime
	\end{align}
	\noindent noting that one can substitute the averaging of a set of damped responses by a temporal averaging performed on the duration of a single acquisition. The envelope of each vector, expressed on a logarithmic scale, is then determined with this technique to extract the associated spread times:
	\begin{align}
	A_n(t) = 10 \log_{10} \frac{ \int_{t}^{+\infty} \lvert \mathfrak{u}_n(t^\prime)\lvert ^2 dt^\prime }{ \int_{0}^{+\infty} \lvert \mathfrak{u}_n(t)\lvert^2 dt }
	\end{align}
	It is then possible for each vector $\mathfrak{u}_n(t)$ to determine an associated spread time $\tau_n$ by linear regression, considering around the centroid time $\bar{t}_n$ (Fig.~\ref{fig:centroid_spread}). The extraction of these parameters allows the quantification of the spread of each subspace, reflecting the progressive decrease of the temporal coherence within this environment. The analysis window being determined by the frequency sampling considered for these measurements, these approaches remain nevertheless limited by the time truncation of the vectors of highest indices. The spread of singular vectors, as observed in Fig.~\ref{fig:SVD}, is indeed limited beyond half of the indices by the frequency sampling $d\nu \approx 7.5\text{MHz}$, defining the maximum acquisition time $T_{\text{max}} = 1/d\nu \approx 133\text{ns}$. \textcolor{green}{It will thus be necessary to keep a critical eye on the processing carried out on the time vectors impacted by this truncation, represented by a red gradient on Figs.~\ref{fig:centroid_spread} and \ref{fig:utl1}}. The definition of a frequency resolution appropriate to the decay time of the cavity under study ensures that the singular vectors beyond this threshold contribute only to one hundredth of that of the most significant subspaces. The use of linear regressions also implies a certain number of limitations linked to the determination of short reverberation times as well as uncertainties directly visible through the fluctuations of the values of $\tau_n$. As an alternative, it is proposed to consider the $\ell_1$ norm for the determination of a temporal coherence metric of each vector, stating that the latter will necessarily increase with the spread of the considered responses~(Fig.~\ref{fig:utl1}).
	
\begin{figure}[ht]
		\centering
		\includegraphics[width=1\linewidth]{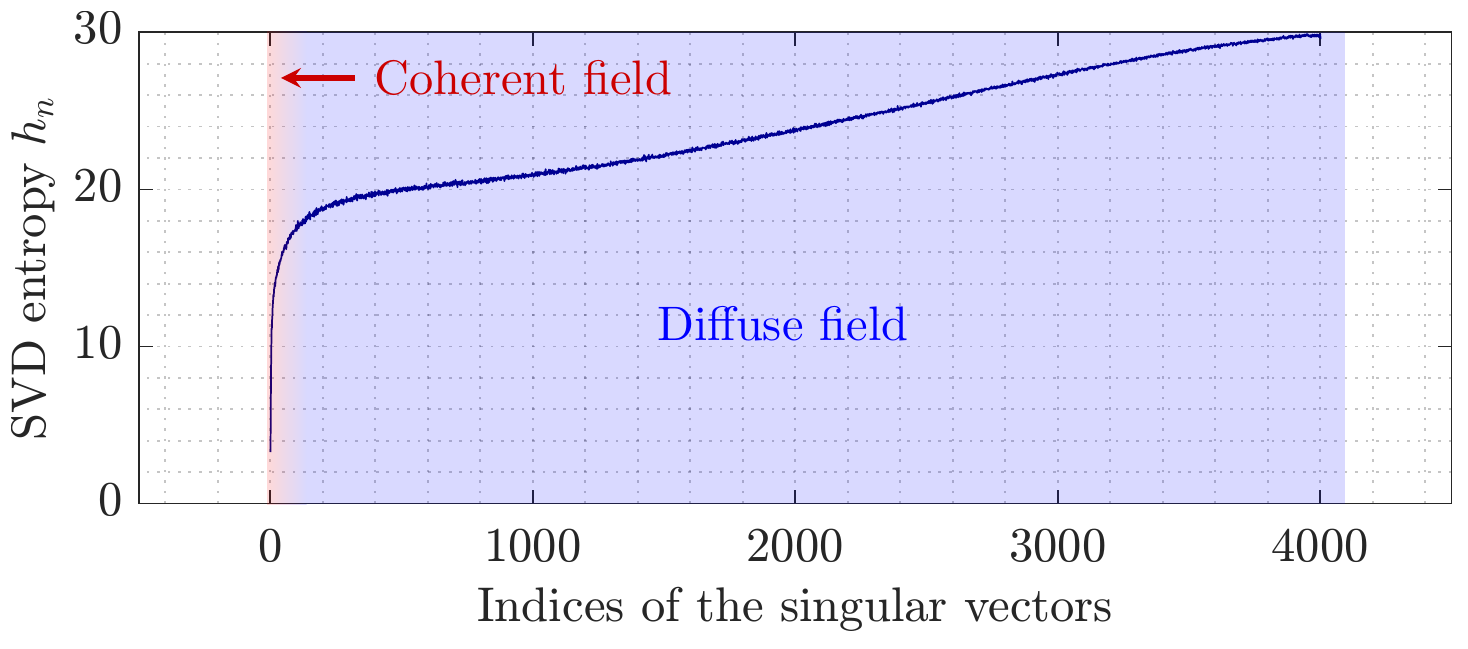}
		\caption{SVD entropy $h_n$ computed for the set of matrices~$\mathbf{V}_n$.}
		\label{fig:v_svd_entropy}
	\end{figure}

	The $\ell_1$ norm of the $\mathbf{\mathfrak{u}}_n$ singular vectors more clearly reveals a rapid spreading of the temporal responses from the first singular values, corresponding to the conversion from quasi-ballistic states to highly reverberating subspaces. The effect of time gating can also be observed, causing a decrease in this metric beyond half of the singular vector indices. Having presented the temporal characteristics of the subspaces describing such fields at the output of complex media, it now seems interesting to study the spatial coherence of these same structures. The processing of singular vectors $\mathbf{v}_n$ must be adapted to the fact that the spatial field distributions are vectorized. As previously introduced in Fig.~\ref{fig:SpatialModes}, it is possible by reshaping the $\mathbf{V}_n$ spatial matrices from the associated $\mathbf{v}_n$ vectors to take into account the interaction of adjacent pixels to determine the level of coherence of these complex spatial distributions. It is proposed in this context to use again the singular value decomposition to determine the level of spatial coherence of each $\mathbf{V}_n$ matrix. The spectrum of singular values determines by its shape the number of independent structures of a matrix, and tends towards a set of identical values in the case of a full rank square matrix where all rows and columns are perfectly uncorrelated. Localized and/or spatially correlated field distributions in phase will thus tend to decrease the number of subspaces needed to describe these field matrices. The SVD entropy $h_n$ seems to be a particularly well suited metric for this work~\cite{varshavsky2006novel}, defined as follows for each matrix~$\mathbf{V}_n$:
	\begin{align}
	h_n = - \sum_i \sigma_{n,i} \log_2(\sigma_{n,i}) 
	\end{align}
	
	\noindent where $\sigma_{n,i}$ corresponds to the singular values of index $i$ for each matrix~$\mathbf{V}_n$. This metric allows the determination of the number of independent structures required to describe the considered spatial matrices, \textcolor{green}{each $\mathbf{V}_n$-matrix being associated to a corresponding value $h_n$}. The evolution of the SVD entropy for all these matrices is given in Fig.~\ref{fig:v_svd_entropy}. 
	
	An inflection point appearing at the first singular values clearly helps identifying the change of propagation regime within this environment. This section will have introduced coherence metrics based on the computation of the $\ell_1$-norm of the temporal vectors and the entropy SVD of the spatial vectors, facilitating the study of the characteristics of the considered fields.\\
	
\begin{figure}[ht]
		\centering
		\includegraphics[width=1\linewidth]{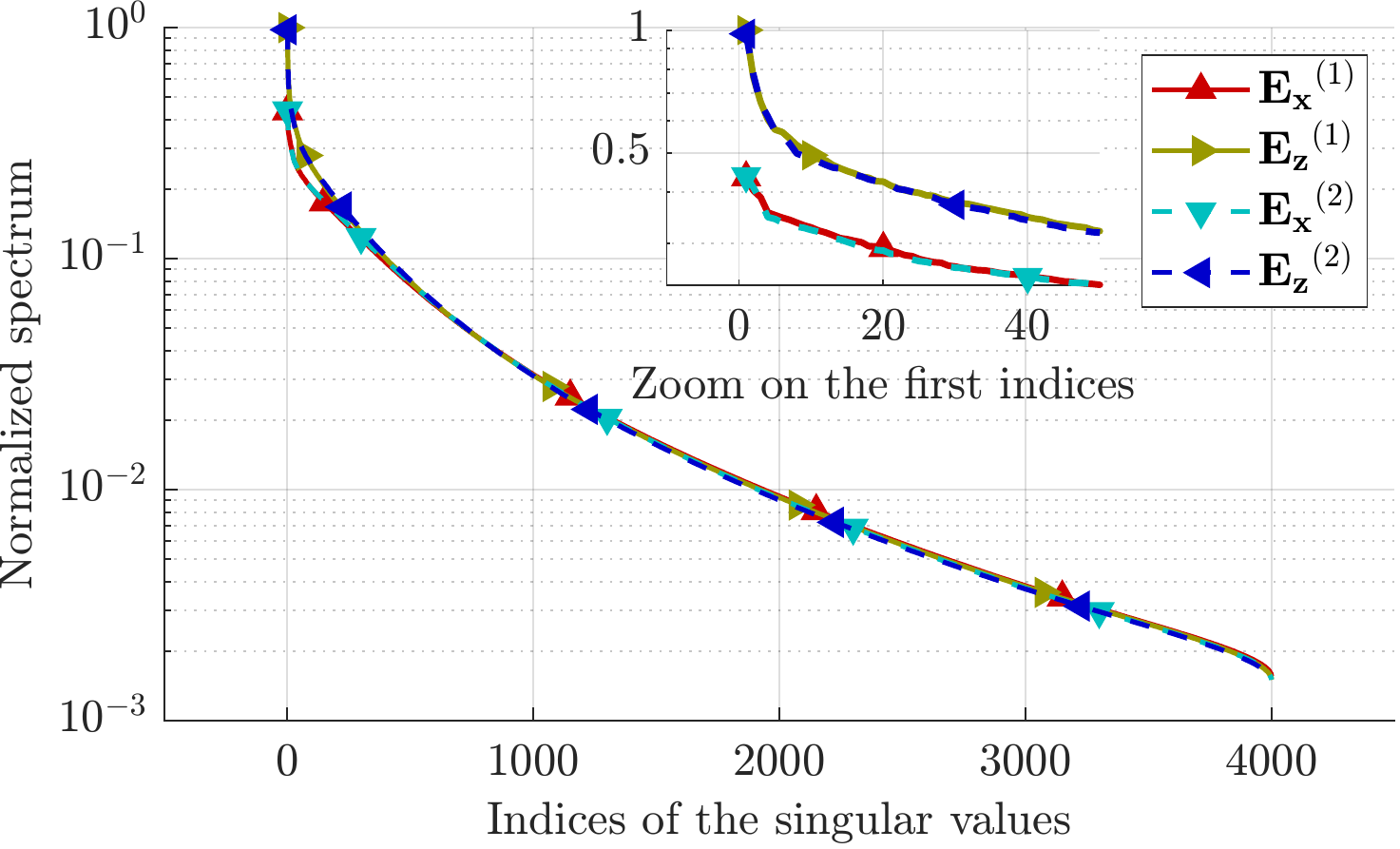}
		\caption{Singular value spectra of the fields $\mathbf{E}_{x}^{(1)}$,$\mathbf{E}_{z}^{(1)}$, $\mathbf{E}_{x}^{(2)}$ and $\mathbf{E}_{z}^{(2)}$. The spectra are normalized by the largest singular value obtained with $\mathbf{E}_{z}^{(1)}$. The cavity ports, whose electric field is oriented along the vertical axis $z$, couple more easily to the measurement probe when the latter is co-polarized.\vspace{-0.5cm}}
		\label{fig:svd_all_polars}
	\end{figure}

	This paper is completed with the excitation of a new cavity port (Fig.~\ref{fig:Cavity}) and by optionally rotating the scanning probe. In such conditions, it is therefore four matrices noted respectively $\mathbf{E}_{x}^{(1)}$, $\mathbf{E}_{z}^{(1)}$, $\mathbf{E}_{x}^{(2)}$ and $\mathbf{E}_{z}^{(2)}$ which can be determined with the excitation of input ports of index 1 and 2 and field measurement along the two transverse field polarization $x$ and $z$ (Fig.~\ref{fig:svd_all_polars}) 
	
\begin{figure}[ht]
		\centering
		\includegraphics[width=\linewidth]{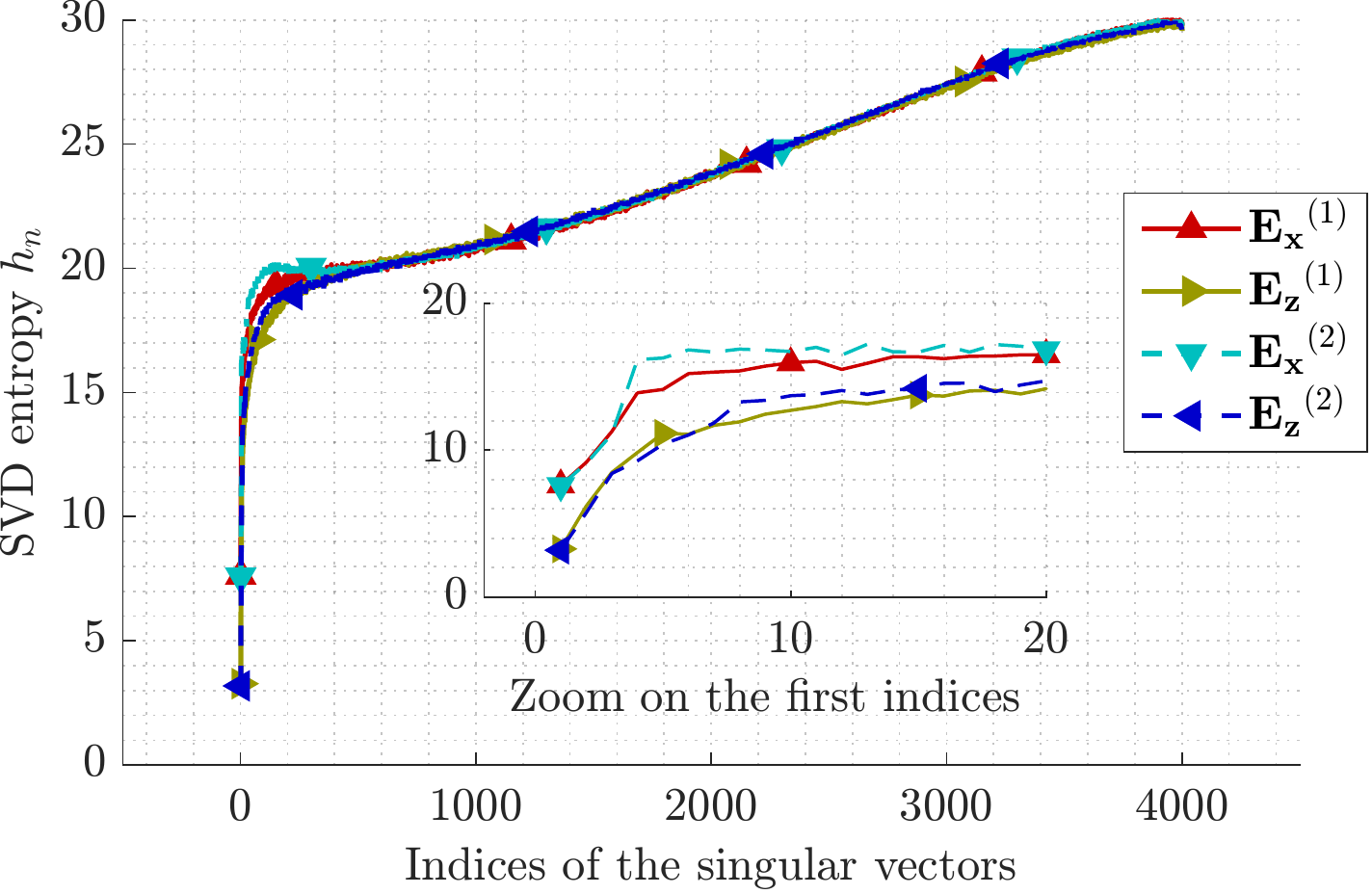}
		\caption{SVD entropy of the $\mathbf{V}_n$ matrices derived from the $\mathbf{E}_{x}^{(1)}$,$\mathbf{E}_{z}^{(1)}$, $\mathbf{E}_{x}^{(2)}$, and $\mathbf{E}_{z}^{(2)}$ fields. The number of spatial structures needed to describe the electric fields depends directly on the polarization conversion of the quasi-ballistic paths. The cross-polarized fields $\mathbf{E}_{x}^{(1)}$ and $\mathbf{E}_{x}^{(2)}$ thus require more orthogonal subspaces to describe their distribution.}
		\label{fig:entropy_all_polars}
	\end{figure}

	The $\mathbf{E}_{z}^{(1)}$ and $\mathbf{E}_{z}^{(2)}$ fields are measured when the scanning probe is co-polarized with the cavity's excitation ports 1 and 2. The amplitude of the first singular values, corresponding the contribution of a direct path, is then a factor of 2 times larger than those of the cross-polarized fields $\mathbf{E}_{x}^{(1)}$ and $\mathbf{E}_{x}^{(2)}$, reflecting the impact of early polarization conversion that occurs in the cavity. Following the previous analysis, it is possible to anticipate an impact on the spatial structuring of the first singular vectors $\mathbf{v}_n$ associated with each measured near field. We thus compare the SVD entropy for each case, considering again the $\mathbf{V}_n$ matrices obtained by reshaping the $\mathbf{v}_n$ vectors (Fig.~\ref{fig:entropy_all_polars}).
	
\begin{figure}[ht]
		\centering
		\includegraphics[width=\linewidth]{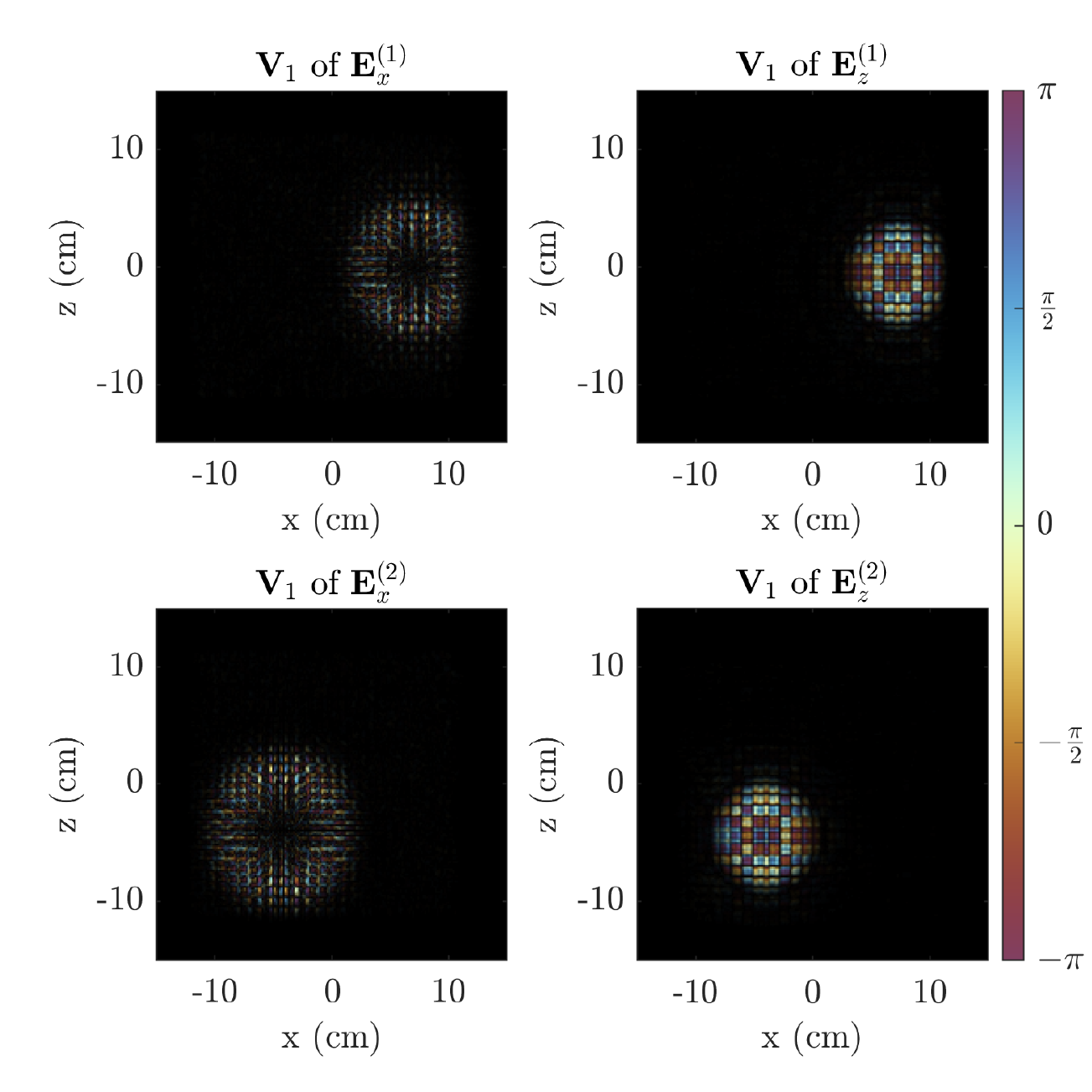}\\[-0.4cm]
		\caption{Matrices obtained by de-vectorizing the singular vectors $\mathbf{v}_1$ derived from the four studied electric fields studied. The action of the polarization conversion by diffraction on the cavity irises is clearly visible in the first singular vectors associated with $x$-polarized fields.}
		\label{fig:V1_all_polars}
	\end{figure}

	The effect of polarization conversion is again clearly visible in this analysis. The first singular vectors require twice as many independent spatial structures when the direct coupling is measured through polarization conversion, reflecting a reduced spatial coherence. After the inflection point corresponding to the change of propagation regime within the cavity, it is interesting to note that the entropy of the singular vectors tends towards an identical distribution, whatever the excitation port and the polarization of the measured electric field. \textcolor{green}{The relative difference between the smallest and largest value of $h_n$ determined with our four electric field distributions is thus of the order of 1\% beyond half of the indices of singular values}. Over long timescales, the characteristics of the spatial field distributions do not depend on the initial excitation conditions and seem dominated by the geometry of the considered environment. The impact of diffraction on the first singular vectors can be illustrated more directly by representing the $\mathbf{V}_1$ matrices associated to each field distribution (Fig.~\ref{fig:V1_all_polars}).
	
\begin{figure}[ht]
		\centering
		\includegraphics[width=\linewidth]{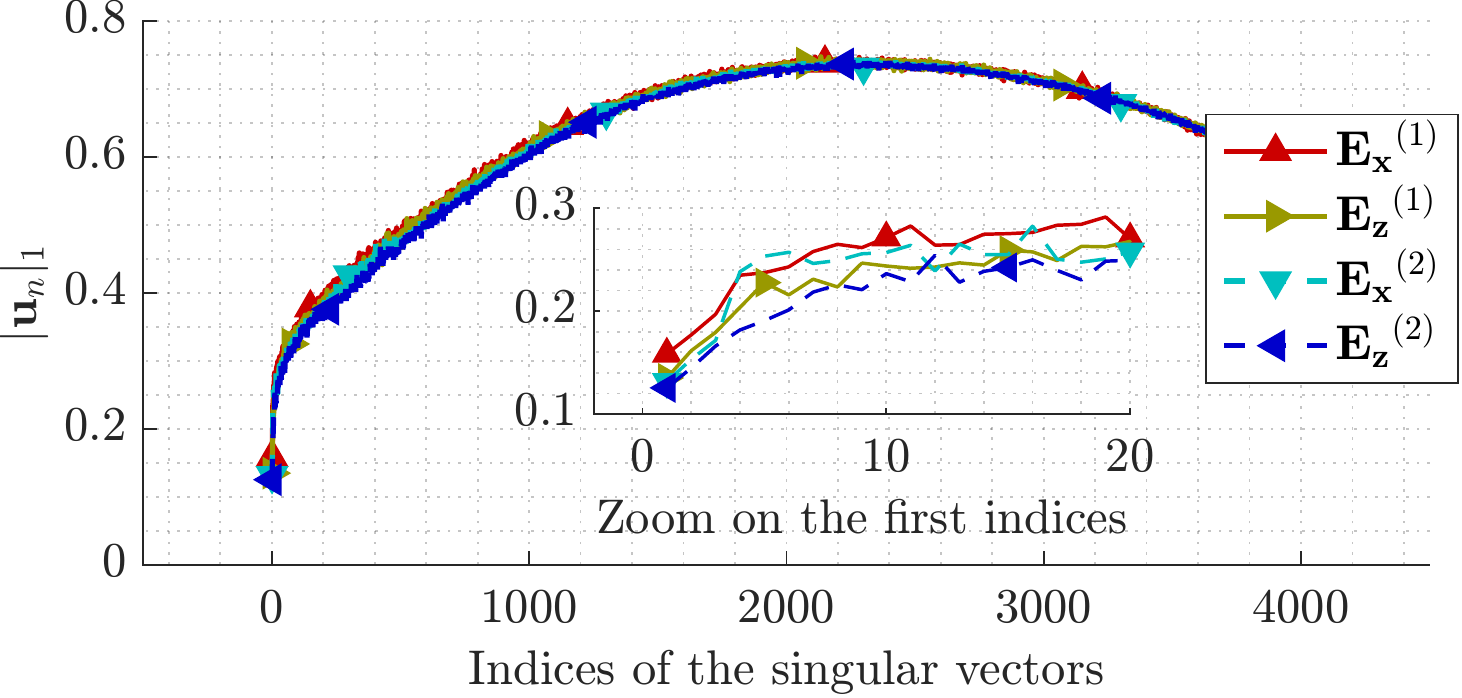}
		\caption{Evolution of the temporal coherence of the singular vectors $\mathbf{u}_n$, determined through their norm $\ell_1$ for the four electric fields considered.}
		\label{fig:utl1_all_polars}
	\end{figure}

	The extraction of these dominant spatial structures in the description of the electric fields allows to highlight the nature of the ballistic paths in each considered case. The spatial distributions of the cross-polarized cases ($\mathbf{E}_x^{(1)}$ and $\mathbf{E}_x^{(2)}$) can be used to guess the mechanism of polarization conversion by diffraction. The projection of the electric field along the longitudinal dimension $y$ associated with the excitation of oblique incidence wave vectors then undergoes a conversion along the two transverse dimensions, justifying the absence of field in the axis of the excitation ports. These tools help highlighting the level of spatial coherence of the main structures characterizing the fields radiated at the output of this reverberant medium. This study is finally completed by an analysis of the temporal characteristics associated with the four electric fields analyzed (Fig.~\ref{fig:utl1_all_polars}).
	
	The computation of $\vert \mathbf{u}_n \lvert_1$ for all the considered cases highlights the weak impact of the polarization conversions and of the position of the excitation ports on the temporal spread of the singular vectors, confirming that spatial coherence characteristics can be affected without major impact on the associated temporal coherence metrics.
	
	This section has introduced a number of useful methods for the characterization of a field propagated through a complex medium, decomposing the latter into subspaces ordered by level of contribution and coherence. In the next section of this paper, an application of this work is proposed in the context of source localization experiments using frequency-diverse and spatially diffuse fields. 
	
	\section{Application to source localization by frequency-diverse diffuse field radiation}
	
	In this paper, numerical techniques are studied to characterize singular vectors describing the temporal and spatial evolution of fields in complex media. Following the identification of the coherence level of each subspace, it is now possible to filter those which are the most favorable to the targeted applications. We anticipate, for example, a possible application of this work to the characterization of antenna radiation in environments where the control of propagation channels cannot be simply ensured with absorbing materials. It would then be possible to limit the impact of weakly coherent subspaces, preserving the structures associated with the direct path. In an opposite context, it is sometimes desirable to suppress the temporally coherent structures of radiated fields. An application example is thus proposed in a context of computational imaging performed with the same cavity~\cite{fromenteze2015computational,tondo2017computational,imani2020review}. The field probe initially used for the radiated near field measurement is now placed at a longer distance from the radiating plane. Therefore, our objective is to locate the source from the sole measurement of the transfer function between the latter and one of the ports of the cavity. The success of such an operation relies initially on the computation of the field radiated in space, discretized into a certain number of voxels according to the resolution limits imposed by the dimensions of the radiating aperture and by the operating bandwidth~\cite{fromenteze2016phaseless}. If the electric field expressed in the frequency domain at the source location is sufficiently different from all other positions in the region of interest, a simple correlation calculation then allows its localization. In such conditions, the optimization of the decoherence of the radiated fields minimizes the possible similarities of the frequency distributions computed at the different locations, thus limiting the ambiguity associated with the determination of the source position.\\
	
	The radiated electric field is computed by propagation of the near field measured by feeding the first excitation port of the cavity. The propagation of the near-field scans performed for the localization experiments presented in this work is carried out using the discrete dipole approximation~\cite{jackson1999classical,lipworth2015comprehensive,caloz2020electromagnetic}. The measured tangential electric field vector $\mathbf{E}_{\text{tan}}$ is thus transformed into a set of point sources, considering that the spatial mesh is chosen fine enough so that the field is constant at the scale of each cell. We thus determine for each elementary surface of area $da$ the electric $\mathbf{p}$ and magnetic $\mathbf{m}$ dipoles according to the following relations:  
	\begin{align}
	&\mathbf{m} = \frac{2}{i 2 \pi \nu \mu_0} \int \hat{\mathbf{n}} \times \mathbf{E}_{\text{tan}}\, da \label{eq:m}\\
	&\mathbf{p} = \epsilon_0 \hat{\mathbf{n}} \int \hat{\mathbf{n}} \cdot \mathbf{E}_{\text{tan}}\, da
	\end{align}
	
	The vector $\hat{\mathbf{n}}$ refers to the normal to the surface on which the tangential field $\mathbf{E}_\text{tan}$ is determined, i.e. $\hat{\mathbf{y}}$ in the reference frame chosen for this work. The constants $\mu_0$ and $\epsilon_0$ correspond respectively to the magnetic permeability and the dielectric permittivity of vacuum. The dipoles can then facilitate the determination of the electric $\mathbf{E}$ field radiated in space:
	\begin{align}
	\mathbf{E} = Z_0 \mathbf{G}^\text{me} \mathbf{m} + \mathbf{G}^\text{ee} \mathbf{p}
	\end{align}
	
	\noindent where the $\mathbf{G}$ matrices are the dyadic Green's functions propagating the different electric and magnetic quantities. The contribution of the electric dipoles $\mathbf{p}$ can be neglected in the context of this work, considering that the scalar product $\hat{\mathbf{n}} \cdot \mathbf{E}_{\text{tan}}$ tends to zero if the normal to the characterization plane is orthogonal to the polarization axes of the transverse electric field. In practice, the various surface irregularities (screws, iris milling) can induce the generation of effective electric dipoles, but their contribution along the axis of propagation remains negligible compared to that of the magnetic sources on the entire characterized surface. The expression of the radiated electric field then takes the following form~\cite{lipworth2015comprehensive,fromenteze2017computational}:
	\begin{align}
	\mathbf{E}(\mathbf{r}_j) = -Z_0\, k^2 \sum_i (\hat{\mathbf{r}}_{ij} \times \mathbf{m}_i)\, \frac{e^{-i k \mathbf{r}_{ij}}}{4 \pi \mathbf{r}_{ij}}\, \left(1-\frac{i}{k\,\mathbf{r}_{ij}}\right) \label{eq:Propag}
	\end{align}
	
	The sum of the contributions of magnetic dipoles of index $i$ then allows the determination of the field at each position $\mathbf{r}_j$, considering Euclidean distances $\mathbf{r}_{ij} = \lVert \mathbf{r}_i - \mathbf{r}_j \lVert$ and unit vectors $\hat{\mathbf{r}}_{ij} = \frac{ \mathbf{r}_i - \mathbf{r}_j}{\lVert \mathbf{r}_i - \mathbf{r}_j\lVert}$. The wavenumber $k = 2\pi \nu /c$ is defined for all the frequency samples of the operating bandwidth, set between 70GHz and 100 GHz. We recall that the localization experiments performed in this work require the computation of the radiated field when the first port of the cavity is fed. This field is determined at a distance $y = R =  25 \text{cm}$ from the front panel of the cavity, on a surface $\mathbf{r}^\prime$ of  $15\text{cm}\times 15\text{cm}$. Each axis is sampled by 51 points, i.e. $n_{r^\prime} = 2601$. The diffraction limit is defined at the highest frequency (100GHz), determining the smallest radiated wavelength $\lambda_{\text{min}}$. Considering a distance $R = 25\text{cm}$ and a square radiating aperture of $L_{x,z}=220\text{mm}$ side, the Rayleigh diffraction limit $\delta_{x,z}$ is calculated as follows:
	\begin{align}
	\delta_{x,z} = \frac{\lambda_{\text{min}} R}{L_{x,z}} \approx 3.4\text{mm}
	\end{align}	
	
	The imaged space $\mathbf{r}^{\prime}$ is defined by pixel size of 2.5mm, chosen slightly oversampled compared to the resolution limits in order to mitigate the effects of grid shift between the target position and the nearest pixels.
	
\begin{figure}[ht]
		\centering
		\includegraphics[width=\linewidth]{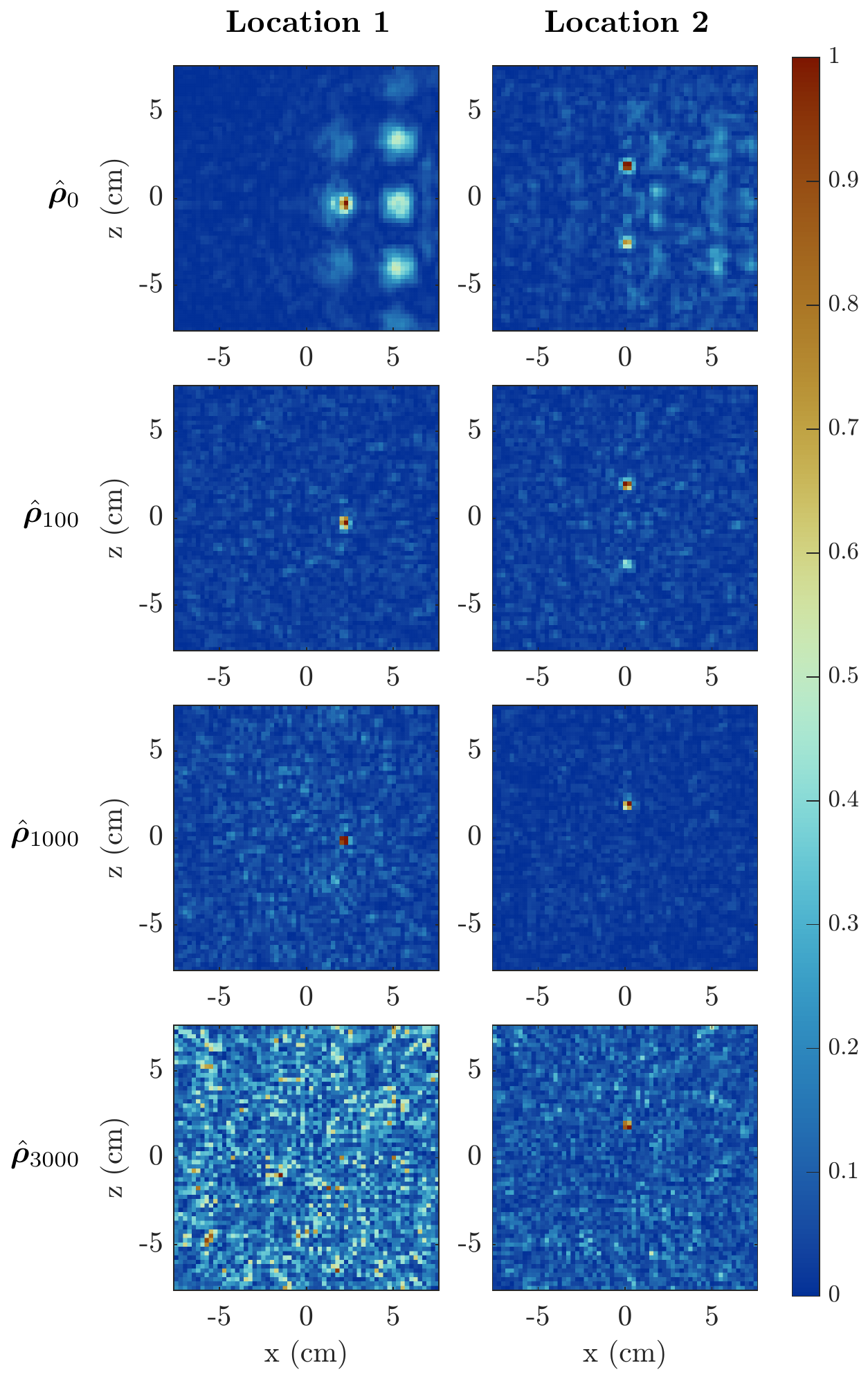}
		\caption{Localization experiments of radiating sources set at $(x,z) = (2\text{cm},0\text{cm})$ and $(x,z) = (0\text{cm},2\text{cm})$, illustrating the deleterious impact of the most coherent subspaces on the ambiguity of the reconstructed positions.}
		\label{fig:Loc}
	\end{figure}

	In the absence of filtering of the subspaces according to their level of coherence, we first determine the field matrix $\mathbf{E}_0 \in \mathbb{C}^{n_{\nu} \times n_{r^\prime}}$. The singular value decomposition filtering is then performed directly on the tangential electric field $\mathbf{E}_{\text{tan}}$ measured by the scanner. Using the notations introduced in the paper, the data are arranged to form two scalar near-field matrices $\mathbf{E}_{x}^{(1)}$ and $\mathbf{E}_{z}^{(1)}$, both complex and of dimensions $n_\nu \times n_r$. Each matrix $\mathbf{E}_{x,z}^{[m]}$ is then diagonalized using a singular value decomposition, in order to remove the $m$ first most significant and coherent components :
	\begin{align}
	\mathbf{E}_{x,z}^{[m]}= \sum_{n=m+1}^{n_\nu} \sigma_n \mathbf{u}_n \mathbf{v}_n^\dagger \label{eq:Field}
	\end{align}
	
	\noindent where the singular vectors $\mathbf{u}_n$ and $\mathbf{v}_n$ correspond to the polarizations $x$ and $z$ considered. Following the suppression of the most coherent subspaces, the filtered tangential fields allow the determination of point sources enabling the propagation of electric fields following Eqs.~(\ref{eq:m}) to   (\ref{eq:Field}). In the context of this study, we are specifically focused on the field matrix \mbox{$\mathbf{E}_m \in \mathbb{C}^{n_\nu \times n_{r^\prime}}$} determined in the imaged space $\mathbf{r}^\prime$ according to the polarization $z$ to match to that of the source. The measurements of transfer functions $s^{(q)}(\nu)$ are conducted for two source positions of index $q$, successively arranged at $(x,z) = (2\text{cm},0\text{cm})$ and $(x,z) = (0\text{cm},2\text{cm})$ in the plane $y = 25 \text{cm}$. In the context of a simplified scalar model, the emissivity of the source $\rho^{(q)}(r^\prime)$, considered independent of the frequency on the analysis band, is then related to the measured signals according to the relation $\mathbf{s}^{(q)} = \mathbf{E}_0\, \bm{\rho}^{(q)}$. The emissivity can then be estimated by compensating the phase of the field distributions in space and frequency, thus providing a spatial correlation of the measured signal with the pre-determined field:
	\begin{align}
	\hat{\bm{\rho}}_{m}^{(q)} &= \mathbf{E}_m^\dagger\, \mathbf{s}^{(q)} =  \mathbf{E}_m^\dagger \mathbf{E}_0\, \bm{\rho}^{(q)} \label{eq:recLoc}
	\end{align}
	The results obtained for different filtering indices $m$ are presented in Fig.~\ref{fig:Loc}. This demonstration highlights the effect of filtering the first indices to remove the influence of the most direct and coherent field paths in time and space. It is even possible for the $\hat\rho_{0}^{(1)}$ reconstruction in the first source position to identify quite distinctly the diffraction pattern caused by the radiation from the excitation port through the iris grid. While a single source is arranged facing the cavity in the case of the second localization experiment, an artifact is clearly visible in $\hat\rho_0^{(2)}$ and $\hat\rho_{100}^{(2)}$, at a position symmetric with respect to the $z=0$ axis. The latter appeared for several source positions in our experiments and seems to be related to the geometry of the cavity, still presenting symmetries despite the presence of a portion of metal sphere at one of its corners. The progressive increase of the filtering index $m$ allows nevertheless to remove these artifacts. Considering that higher indices are associated with longer mean propagation times (Fig.~\ref{fig:centroid_spread}), it seems intuitive that the associated spatial field distributions are increasingly exposed to the impact of the convex boundary condition imposed in this environment, limiting the ambiguity related to the response of this artifact with respect to the real position of the source. Such properties are studied in more detail in the context of time-reversal experiments, where the ergodic properties of the media considered ensure that all locations in an environment will be visited by incident waves over large propagation times~\cite{draeger1997one}. 
	
	The progressive increase of the filtering index $m$ does not allow to continuously improve the quality of the reconstructed images. It is indeed necessary to consider the effect of the level differences of the singular values associated with these subspaces, of which the example $m=3000$ illustrated in Fig.~\ref{fig:Loc} corresponds to amplitudes more than 100 times lower than the first singular values (Fig.~\ref{fig:svd_all_polars}). In the presence of additive noise inherent to any measurement, the latter are indeed the most likely to be corrupted, justifying the development of pseudo-inversion by regularization based on the suppression of the highest indices~\cite{hansen1990truncated,cannon2010singular}.\\
	

	\color{black}
	
	This analysis is finally completed by a remark. The monotonic decay of the spatial and temporal coherence of the considered subspaces implies a possible substitution of the detailed processing by a simple time gating applied directly to the near field matrix. In such conditions, the cancellation of the first temporal samples limits the impact of the strongly correlated subspaces according to the spatial and frequency dimensions. We thus define a time gating matrix $\mathbf{T}_m = \text{diag}(\textbf{q}_m)$, where $\text{diag}(\textbf{q}_m)$ is a vector composed of zeros except for the samples of index $m+1$ to $n_\nu$ which are set to 1. A temporal filtering of the first $m$ indices is applied on the near field matrix $\mathbf{E}$ as follows :
	\begin{align}
		\mathbf{E}_{tg-m} &= \mathbf{W}\,\mathbf{T}_m \mathbf{W}^{\dagger}\mathbf{E}\\
						  &= \mathbf{W}\,\mathbf{T}_m \mathbf{W}^{\dagger} \mathbf{U} \mathbf{S} \mathbf{V}^\dagger\\
						  &= \mathbf{W}\,\mathbf{T}_m \mathbf{U_t} \mathbf{S} \mathbf{V}^\dagger
	\end{align}

\begin{figure}[ht]
		\centering
		\includegraphics[width=\linewidth]{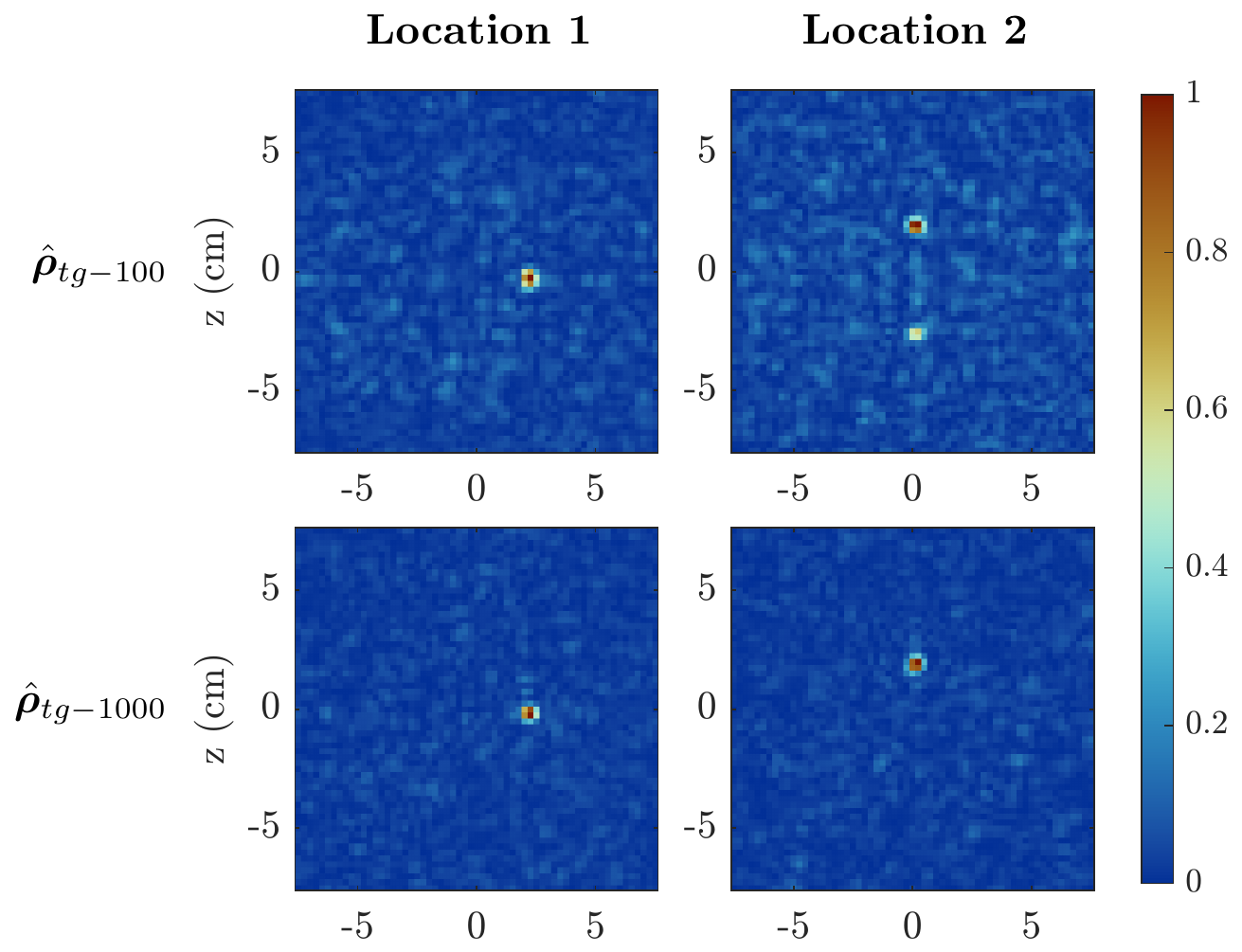}
		\caption{Source localization achieved with time gated near fields.}
		\label{fig:Loc_tg}
	\end{figure}

	This decomposition allows us to highlight that the application of filtering by $\mathbf{T}_m$ is performed indirectly on the singular vectors of the $\mathbf{U_t}$ matrix. The diagonal form of the latter (Fig.~\ref{fig:Ut}) implies that a suppression of the $m$ first temporal samples will have an impact comparable to the suppression of the $m$ first singular values. The near field $\mathbf{E}_{tg-m}$ is thus propagated using equations (\ref{eq:m}) and (\ref{eq:Propag}) and used for the estimation of the position of the sources $\bm{\rho}_{tg-m}$ following Eq.~(\ref{eq:recLoc}). The results for $m = 100$ ($t = 30$~ns) and $m = 1000$ ($t = 300$~ns) are given in Fig.~\ref{fig:Loc_tg}. 
	
	Benefiting from the previous analyses, image reconstruction can be greatly accelerated by substituting the singular value decomposition of the near field by a succession of direct and inverse Fourier transforms, ideally performed with fast Fourier transforms, and simple temporal filtering. It is nevertheless necessary to maintain a critical eye on the use of such a method, based on a monotonic decrease of the temporal and spatial coherence of the field during its propagation. As an example, the Saleh-Venenzuela model in the field of indoor and broadband telecommunications describes stochastically the impulse responses between transmitter and receiver formed by a certain number of temporal clusters constituting a succession of damped responses separable in time~\cite{saleh1987statistical,meijerink2014physical}. The study of scattering along quasi one-dimensional media similarly allows the identification of configurations where the contribution of ballistic paths is much smaller than that of more complex paths~\cite{davy2015transmission}, implying a potentially more complex mapping of coherence states as a function of the singular value index. A deeper analysis of the spatial and temporal coherence thus allows in all situations and with the help of the previously introduced metrics to identify and select the most suitable subspaces for the considered applications.
	
	\color{black}

	\section{Conclusion}
	
	\textcolor{blue}{In conclusion, this study has allowed the development of techniques for the decomposition of electromagnetic fields propagated through a complex medium, discriminating orthogonal structures composing the latter according to their level of spatial and temporal coherence. Many references in the scientific literature focus on techniques involving space-space transfer matrices for the calculation and analysis of Wigner-Smith operators. In a similar context, this work has investigated the decomposition of space-frequency or space-time transfer matrices, paving the way for an in-depth analysis of the temporal spreading phenomena occurring during interaction with a complex medium.} Beyond the only perspectives of characterization of these challenging propagation environments, these methods can also be exploited in applications whose operation in conditions of partial coherence may be compromised. Such thresholding on the first singular values and vectors of a linear application can be counter-intuitive for scientists initiated to conditioning and regularization problems~\cite{hansen1990truncated}. The most significant subspaces are by definition the most immune to noise, limiting the amplification of the most diffuse structures under poor signal-to-noise conditions. A thresholding of the strongest and weakest singular values could thus be considered in more disturbed operating conditions. Previous analyses have also shown that the position and polarization of the excitation port within these environments could have a direct effect on the coherence of the most significant orthogonal structures. Coupled with numerical tools, the optimization of applications requiring the use of diffuse fields must therefore necessarily be carried out by limiting as much as possible the direct interaction between the excitation ports of the environments considered and the analysis plane. Finally, we anticipate possible applications of this work to the shaping of electromagnetic fields, exploiting the many degrees of freedom offered by complex media and allowing the local focusing or cancellation of energy with linear combinations of pre-selected subspaces.
	
	\section*{ACKNOWLEDGMENTS}
	 T.~Fromenteze acknowledges the support of the French Agence Nationale de la Recherche (ANR) under reference ANR-21-JCJC-0027-01. The work of O. Yurduseven was supported by a research grant from the Leverhulme Trust under the Research Leadership Award RL-2019-019.
	
	\bibliography{BibBallisticPCA}

\end{document}